\documentclass[useAMS,usenatbib]{mn2e}

\usepackage{float} 
\usepackage{epsfig}
\usepackage{graphicx}
\usepackage[latin1]{inputenc}
\usepackage{latexsym}

\def\simlt{\ \raise -2.truept\hbox{\rlap{\hbox{$\sim$}}\raise5.truept   %
\hbox{$<$}\ }}
\def\simgt{\ \raise -2.truept\hbox{\rlap{\hbox{$\sim$}}\raise5.truept   %
\hbox{$>$}\ }}                                                          %

\def\be{\begin{equation}}
\def\ee{\end{equation}}
\def\newline{\hfil\break}

\def\la{\mathrel{\hbox{\rlap{\hbox{\lower4pt\hbox{$\sim$}}}\hbox{$<$}}}}
\def\ga{\mathrel{\hbox{\rlap{\hbox{\lower4pt\hbox{$\sim$}}}\hbox{$>$}}}}

\title[Discovery of a GRG with KAT-7]{Discovery of a suspected giant radio galaxy with the KAT-7 array}

\author[S. Colafrancesco et al.]
{S. Colafrancesco$^{1}$\thanks{E-mail:sergio.colafrancesco@wits.ac.za}, N. Mhlahlo$^{1}$, T. Jarrett$^{2}$, N. Oozeer$^{3,4,5}$ and P. Marchegiani$^{1}$\\
$^{1}$School of Physics, University of the Witwatersrand, Private Bag 3, 2050-Johannesburg, South Africa\\
$^{2}$Astronomy Department, University of Cape Town, Private BagX3, Rondebosch 7701, South Africa\\
$^{3}$SKA South Africa, The Park, Park Road, Pinelands, Cape Town 7405, South Africa\\
$^4$ African Institute for Mathematical Sciences, 6-8 Melrose Road, Muizenberg 7945, South Africa\\
$^5$ Centre for Space Research, North-West University, Potchefstroom 2520, South Africa.}
\begin{document}

\date{Accepted:  Received; in original form:}

\pagerange{\pageref{firstpage}--\pageref{lastpage}} \pubyear{2014}

\maketitle

\label{firstpage}

\begin{abstract}
We detect a new suspected giant radio galaxy (GRG) discovered by KAT-7.   The GRG core is identified with the WISE source J013313.50-130330.5, an extragalactic source based on its infrared colors and consistent with a misaligned AGN-type spectrum at $z\approx 0.3$. The multi-$\nu$ spectral energy distribution (SED) of the object associated to the GRG core shows a synchrotron peak at $\nu \approx 10^{14}$ Hz consistent with the SED of a radio galaxy blazar-like core.
The angular size of the lobes are $\sim 4 ^{\prime}$ for the NW lobe and $\sim 1.2 ^{\prime}$ for the SE lobe, corresponding to projected linear distances of $\sim 1078$ kpc and $\sim 324$ kpc, respectively. The best-fit parameters for the SED of the GRG core and the value of jet boosting parameter $\delta =2$, indicate that the GRG jet has maximum inclination $\theta \approx 30$ deg with respect to the line of sight, a value obtained for $\delta=\Gamma$, while the minimum value of $\theta$ is not constrained due to the degeneracy existing with the value of Lorentz factor $\Gamma$. Given the photometric redshift $z \approx 0.3$, this GRG shows a core luminosity of $P_{1.4 GHz} \approx 5.52 \times 10^{24}$ W Hz$^{-1}$, and a luminosity $P_{1.4 GHz} \approx 1.29 \times 10^{25}$ W Hz$^{-1}$ for the NW lobe and $P_{1.4 GHz} \approx 0.46 \times 10^{25}$ W Hz$^{-1}$ for the SE lobe, consistent with the typical GRG luminosities. The radio lobes show a fractional linear polarization $\approx 9 \%$ consistent with typical values found in other GRG lobes.
\end{abstract}


\section{Introduction}

Extended radio emission in galaxies is associated with both radio jets and lobes and with outflows, seen often as aligned radio sources in the opposite directions with respect to the central compact radio core. 
Giant radio galaxies (GRG) are extreme cases of this phenomenology with jets and lobes extending  on $\sim$ Mpc scales suggesting that they are either very powerful or very old site for electron acceleration. In this respect, GRGs have a crucial role in the acceleration of cosmic rays (CRs) over large cosmic scales (e.g., Kronberg et al. 2004), in the feedback mechanism of AGNs into the intergalactic and intracluster medium (e.g., Subrahmanyan et al. 2008) and in the seeding of large-scale magnetic fields in the universe (e.g., Kronberg et al. 2004) and they are excellent sites to determine the total jet/lobe energetics in AGN-dominated structures (see, e.g., Colafrancesco 2008, Colafrancesco \& Marchegiani 2011).
To date our knowledge of GRGs (see, e.g., Ishwara-Chandra \& Saikia 1999, 2002, Lara et al. 2001, Machalski, Jamrozy \& Zola 2001, Schoenmakers et al., 2001, Kronberg et al. 2004, Saripalli et al. 2005, Malarecki et al. 2013, Butenko et al. 2014) is limited by their sparse numbers and by the difficulty of detecting them over large areas of the sky. Low-frequency radio observations have an enhanced capacity to detect the extended old electron population in these objects (see, e.g., the recent Low Frequency Array (LOFAR) observation of the GRG UGC09555 \footnote{http://www.astron.nl/about-astron/press-public/news/lofar-discovers-new-giant-galaxy-all-sky-survey/lofar-discovers-new-g}), but high-frequency radio observations are less efficient in this task due to the steep-spectra of giant radio lobes. In this context these sources will be ideal targets for the next coming deep, wide-field surveys like, e.g., the ATLAS survey of the Australia Telescope Network Facility (ATNF, see Norris et al. 2009) or the SKA deep surveys that will have the potential to study their population evolution up to high redshifts and thus clarifying their role on the feedback for the evolution of non-thermal processes in large-scale structures.\\ 
In this paper we report the serendipitous discovery of a new extended radio source which is a suspected case of a GRG in the southern hemisphere observed at 1.83 GHz with the seven-dish MeerKAT precursor array, KAT-7,  in South Africa.\\
Throughout the paper a flat, vacuum-dominated Universe with $\Omega_m = 0.32$ and $\Omega_\Lambda = 0.68$ and $ H_0=67.3$ km s$^{-1}$ Mpc$^{-1}$ is assumed.

\section{Observations}
\label{sec:obs}

\begin{figure}
\centering
\includegraphics[width=100mm,height=75mm]{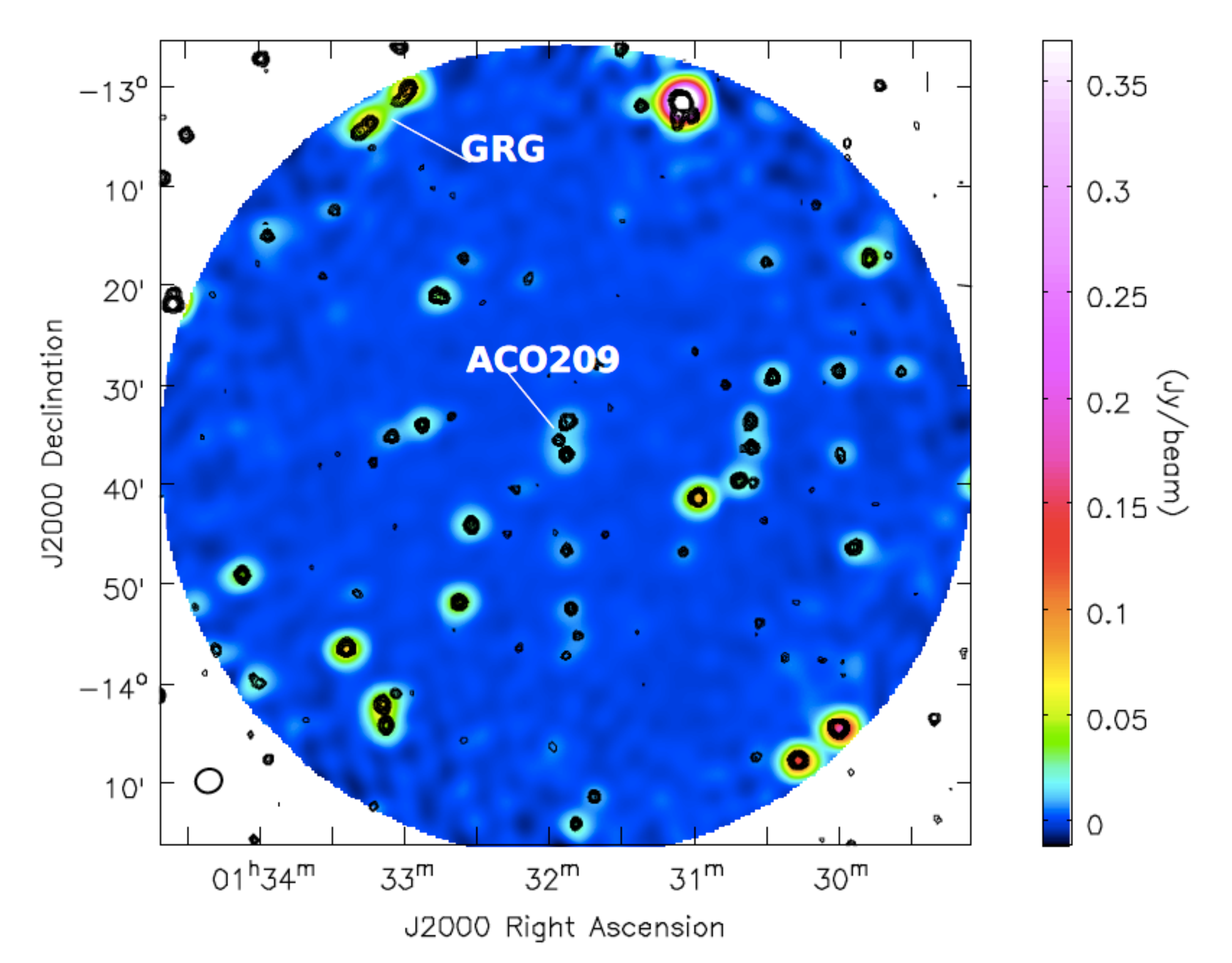}
\caption{The primary-beam corrected KAT-7 image of the ACO209 field observed at 1.83 GHz with KAT-7.
The suspected giant radio galaxy is indicated by the acronym GRG.}
\label{fig:209_nvss}
\end{figure}

The suspected GRG source J013313.50-130330.5 has been detected in the field of the cluster ACO209 (see Fig.\ref{fig:209_nvss}), one of the eight clusters observed with the KAT-7 array at 1.83 GHz during the period 04--11 October 2012 (see Colafrancesco, Mhlahlo \& Oozeer 2015a).\\
The KAT-7 telescope, built as an engineering testbed for the 64-dish Karoo Array Telescope (MeerKAT), has been used to detect HI in nearby galaxies (Carignan et al. 2013), to study extended radio halos in galaxy clusters (Scaife et al. 2015, Riseley et al. 2015) and to study time-variability of radio sources (Armstrong et al. 2013), and we refer to these papers for the technical specifications of the telescope.\\
The ACO209 field was observed with a total bandwidth of 256 MHz divided in 1024 channels. After initial flagging of bad data, only 601 channels with useful data remained.
The National Radio Astronomy Observatory (NRAO) Common Astronomy Software Applications (CASA) package, version 3.4 was used to perform the calibration. Bad data and RFI were removed within CASA. Before calibration, the 601 channels were averaged to 19, each having a width of 12.5 MHz to reduce the size of the data set. The interval width for time averaging was 15s, and 32 channels were averaged to output each of the 19 channels. The choice for these values was done in order to ensure that time/frequency smearing were minimized.\\
For the observation of the ACO209 field, the flux density calibrator source PKS 1934-638 and the phase calibrator source PKS 0117-155 were used to calibrate the data in flux and in phase, respectively. The flux density calibrator source was also used to do bandpass calibration.\\
Information on the observation for the source presented in this work is summarized in Table~\ref{tab:log}.\\
Briggs weighting with a robustness of zero was used in CLEAN to produce the images. After a few steps of self-calibration to reduce residual phase variations and to increase the dynamic range, the primary beam correction was done using task IMMATH in CASA where KAT-7 images were divided by the corresponding primary beam images (i.e. the flux images).
A multi-scale CLEANing algorithm in CLEAN was used in order to detect diffuse, extended structures on both small and large spatial scales.

The KAT-7 field of view around the cluster ACO209 is quite large covering an area of about $\sim 10$ deg$^2$ where several other radio sources are clearly detected (see Fig.\ref{fig:209_nvss}). 
The complete analysis of discrete sources in the field of the ACO209 cluster has been done by Colafrancesco, Mhlalo \& Oozeer (2015b). In order to check the consistency of our KAT-7 image with the NRAO VLA Sky Survey (NVSS) one, we selected 16 bright point-like sources above a threshold flux of 20 mJy in the KAT-7 image (and the same for the NVSS) and we verified that these are coincident with NVSS point-like sources detected above the same flux threshold.

Apart from the suspected GRG we discuss in this paper, we find no other extended radio source in the field of ACO209.

\begin{table*}
\centering
\caption{Details of KAT-7 observation of ACO209}
\label{tab:log}
\begin{tabular}{c c c c c c c c}
\\
\hline\hline
Cluster Name   &  Frequency   &   BW    & Observation Date & Observing time  & Antennas &   FWHM, p.a.   & rms     \\  
	                     &  (MHz)  &  (MHz) &                              &        (hours)       &                  &  ($^{\prime\prime} \times ^{\prime\prime}$), degree     & (mJy/beam)  \\ 
\hline
ACO209	   &  1826.6875   & 234.765  & 04-Oct-2012  & 11.29 & 7  & 159$\times$141, 109  &  1.4              \\
\hline
\end{tabular}
\end{table*}
%

\section{The suspected GRG  J013313.50-130330.5}
\label{sec:rg}

\begin{figure}
\centering
\includegraphics[scale=1.0,width=90mm,height=65mm]{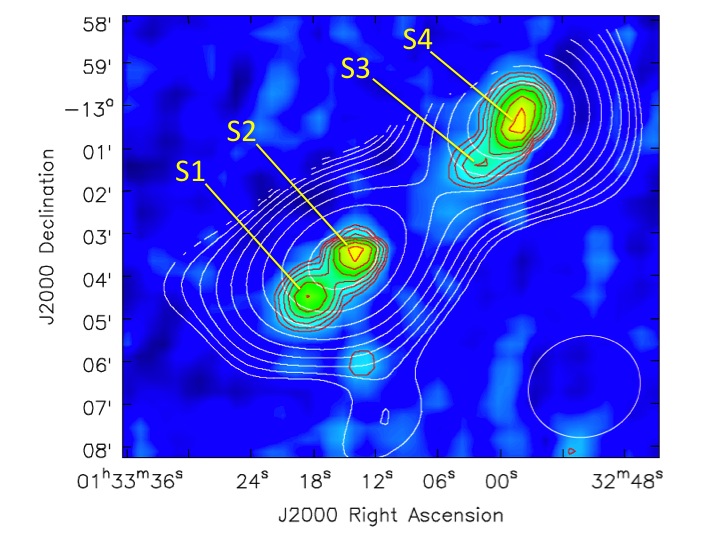}
\caption{A zoom of the primary beam corrected image of the GRG with the KAT-7 contours (white) and the NVSS sources contours (red) superposed.
The restoring beam of the radio image is 159$^{\prime\prime}$ $\times$ 141$^{\prime\prime}$ (PA=109 deg), and the noise level in the image plane is 1.4 mJy/beam. The restoring beam is shown in the bottom-right hand corner. NVSS contours start at 1.6 mJy/beam and then scale by a factor of $\sqrt2$. The NVSS beam is $\sim45^{\prime\prime}$.}
\label{fig:GRG_nvss}
\end{figure}

Fig.\ref{fig:GRG_nvss} shows the NVSS radio contours overlaid onto the primary-beam-corrected KAT-7 image of the region around the object labelled as GRG. 
The NVSS image partially resolves the elongated radio structure in four distinct sources that are labelled as S1 to S4 in Fig.\ref{fig:GRG_nvss}. These NVSS sources have fluxes of $17.6 \pm 1.6$ mJy (S1), $19.5 \pm 1.1$ mJy (S2), $4.1 \pm 0.5$ mJy (S3) and $45.6 \pm 2.4$ mJy (S4) measured at 1.4 GHz. 

The KAT-7 radio telescope array clearly detects an extended emission in the form of two symmetric lobes encompassing the NVSS sources S1 and S2 (the southern-east SE lobe) and S3 and S4 (northern-west NW lobe) suggesting an astrophysical connection with the lobes and the central core of an extended radio galaxy.\\
We first checked the likelihood of the hypothesis that the NVSS sources are associated in two main components, namely S1-S2 and S3-S4, of a double radio source. The empirical link criteria given by  Magliocchetti et al. (1998) provides negative results for this hypothesis yielding an angular separation of 333 arcsec with a link length of 91 arcsec and a flux ratio of 0.74. This result indicates that the extended emission detected by KAT-7 is actually consisting of more than two extended radio sources with a very low likelihood that the NVSS sources S1 and S2 are part of a single extended source.\\
We calculated the flux density of the GRG, from the KAT-7 image at 1.83 GHz, using the Briggs robust 0 weighted image. This gives a lightly better resolution of $\approx 153 \times 144$ arcsec$^2$  with a position angle of 23 degrees. The flux of the SE lobe measured by KAT-7  is $37.5 \pm 3.6$ mJy with a peak flux of $33.0 \pm 3.2$ mJy/beam.
The flux of the NW lobe at the same frequency is $39.0 \pm 2.6$ mJy with a peak flux of  $36.0\pm 3.5$ mJy/beam.\\
Since the spatial resolution of KAT-7 is substantially lower than that of the NVSS we cannot derive a spectral index for the four sources S1 to S4 separately, and this fact also limits our ability to produce spectral index maps of the suspected GRG.\\
However, one can  calculate the spectral index of the NW lobe which encompasses the NVSS sources S3 and S4. Considering the total flux of the NW lobe as measured by KAT-7 at 1.83 GHz and the sum of the NVSS fluxes of the sources S3 and S4 measured at 1.4 GHz, the spectral index (defined so that the flux density writes $S \sim \nu^{-\alpha}$) for the NW lobe (which is the unresolved sum of the sources S3 and S4 in the KAT-7 image) is $\alpha_{(1.4-1.83) GHz} \approx 0.91^{+0.43}_{-0.47}$, consistent with typical spectral indices of GRGs (see, e.g., Lara et al. 2001).

Since the radio image alone cannot discriminate the nature of the extended radio emission and its astrophysical counterpart, we perform in the following a multifrequency analysis of this source in order to identify the candidate central radio galaxy and the nature of the associated radio lobes.

\subsection{Multi-frequency analysis}
\label{sec:results}
\begin{figure}
\centering
\includegraphics[scale=0.6,width=90mm,height=75mm]{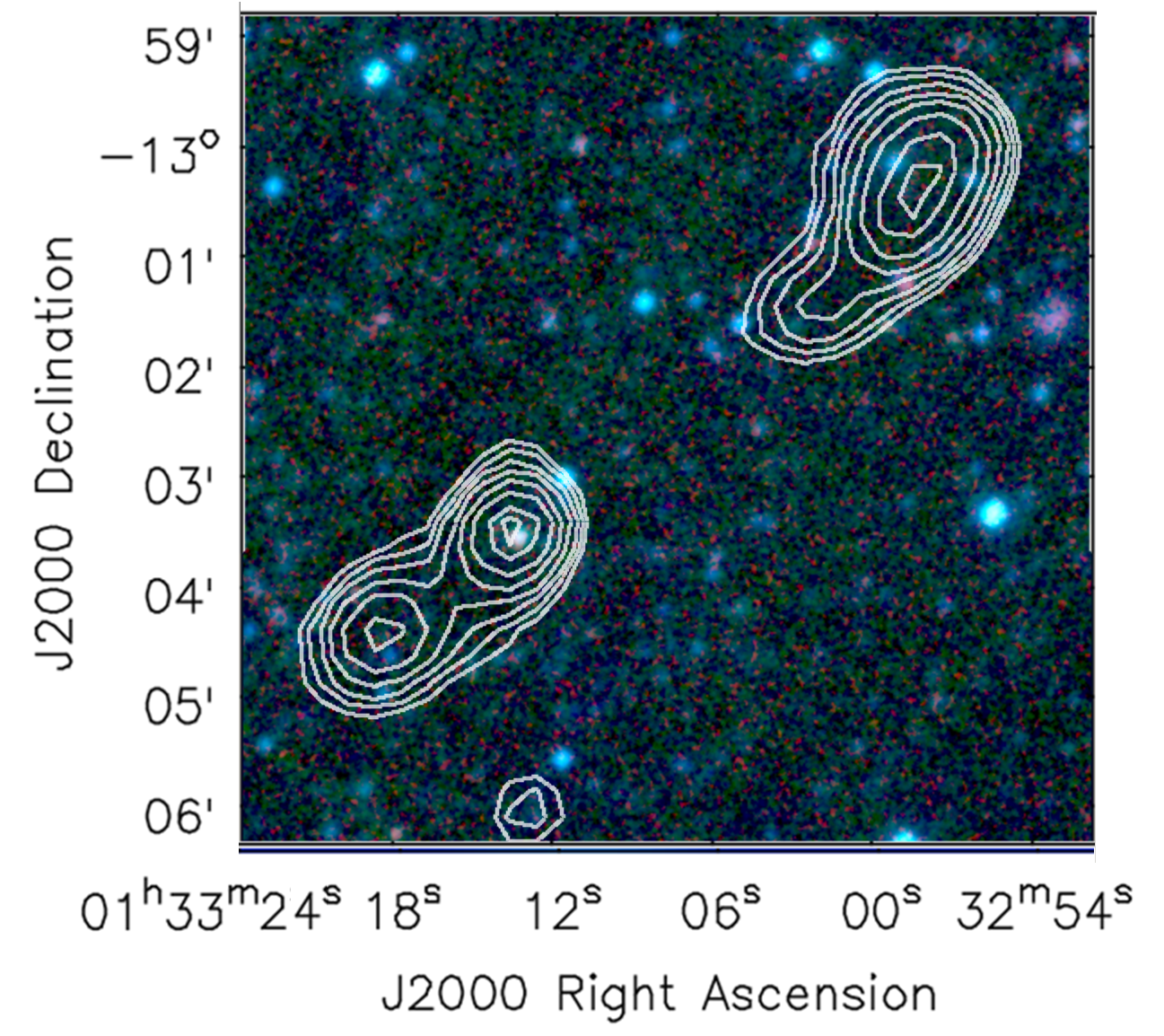}
\caption{The WISE 3-colors, W1 (blue), W2 (green), W3 (red),  image of the field around the GRG with NVSS radio contours as in Fig.\ref{fig:GRG_nvss} superposed.}
\label{fig:209_wise}
\end{figure}

\begin{figure}
\centering
\includegraphics[scale= 0.5,width=80mm,height=80mm]{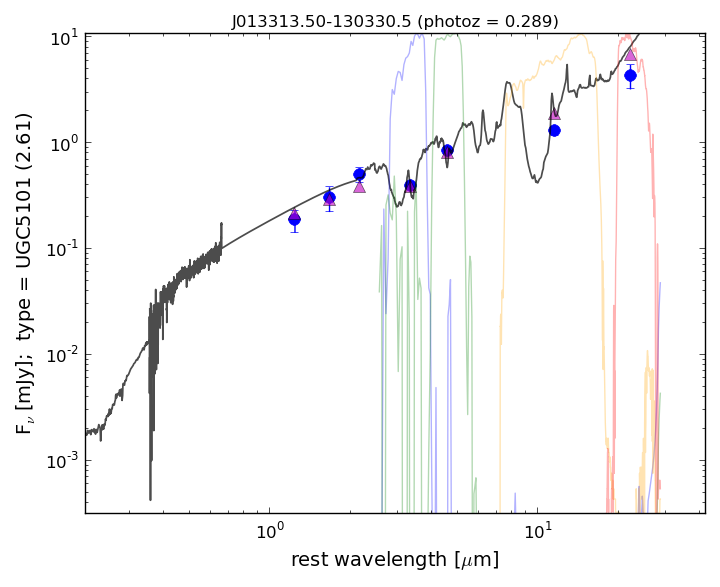}
\caption{The rest-frame SED fitted using the 2MASS and WISE magnitudes: the best fit spectral template is the one of an AGN type from Silva et al. (1998)  at a redshift of 0.289. The blue points are the k-corrected photometry; the magenta triangles are the synthesized photometry based on the UGC5101 template (Brown et al. 2014). The contours denote the WISE filter bandpasses (Jarrett et al 2011).}
\label{fig:209_sed}
\end{figure}

A cross-correlation of the KAT-7, NVSS, infra-red (IR), optical and X-ray sources in the field of the suspected GRG shows that the radio source S2 is associated with the X-ray source 1RXSJ013313.8-13031 in the ROSAT all-sky survey (RASS, Voges et al. 1999) having a count rate of 0.04 cts/s (0.1-2.4 keV band), while none of the other three NVSS sources S1, S3, S4 has an X-ray counterpart.\\ 
At the position of the S2 radio source there is also a well-detected Wide-field Infrared Survey Explorer (WISE) source J013313.50-130330.5. The mid-IR flux densities at 3.4 $\mu$m, 4.6 $\mu$m, 11.6 $\mu$m and 22 $\mu$m are  
$0.71 \pm 0.02$ mJy, $0.89 \pm 0.03$ mJy, $1.49 \pm 0.15$ mJy and $3.45 \pm 1.21$ mJy, respectively. 
The observed  mid-IR colors of this source are consistent with being an unresolved extragalactic source (see Jarrett et al. 2011).  Fig.\ref{fig:209_wise} shows the 3-color image of the same WISE field.  Combining the WISE fluxes with those of near-infrared measurements from Two Micron All-Sky Survey (2MASS),  the resultant spectral energy distribution (SED) exhibits a power-law scaling.  Using the SED templates from Brown et al (2014), the best fit is a spectrum that resembles the one of  an AGN type template from Silva et al. (1998) at a redshift of $\approx 0.3$ (see Fig.\ref{fig:209_sed});  this photometric redshift has been obtained with an uncertainty of $\approx 23\%$ based on the number of WISE bands and on the fit to the available templates.   

The derived photometric redshift yields a realistic indications of the extragalactic nature of this source and the power-law SED provides further hints on the AGN-like nature of the source.\\
This WISE source matches  the radio detection of the source S2  which is the brightest, most compact radio source in the NVSS image of this GRG. 
This indicates that the central radio core S2 with its optical counterpart J013313.50-130330.5 can be the core of the extended radio source which is coincident with the ROSAT X-ray counterpart 1RXSJ013313.8-13031. 
This indicates that the host galaxy seems to be indeed associate with an AGN which is then completely dominating the core radio emission (as indicated by its strong power-law non-thermal spectrum).  

We also point out that there is no WISE or optical source associated with the NVSS sources  S3 and S4 in the NW lobe, while there is an optical source at $\approx 3 ^{\prime \prime}$ from the center of the radio source S1 but with no redshift information and likely a foreground object. 
The absence of any optical or WISE source at the radio mid-point of the sources S1 and S2 further excludes that the SE lobe is a separate radio galaxy. 
The absence of any WISE or optical source at the radio mid-point of the NVSS and KAT-7 image in Fig.\ref{fig:209_nvss} also excludes the possibility that the sources S1-S2 and S3-S4 are the two lobes of a radio galaxy.
We also notice that the NW lobe seems to be in a much less dense environment as compared to the SE lobe. 
Therefore, we conclude that this source is a suspected GRG where the source S2 is coincident with the central galaxy which is also appearing to be associated with the compact core of the radio galaxy, while the source S1 is associated with the SE radio lobe and the sources S3 and S4 are part of the NW radio lobe.\\
The angular size of the GRG lobes is $\sim 4 ^{\prime}$ for the NW lobe and $\sim 1.2 ^{\prime}$ for the SE lobe, corresponding to projected linear distances of $\sim 1078$ kpc and $\sim 324$ for the NW and SE lobes, respectively, assuming a photo-$z$ distance of $\sim 1500$ Mpc.

\begin{figure}
\begin{center}
\hskip-2.2truecm
\includegraphics[width=110mm,height=80mm]{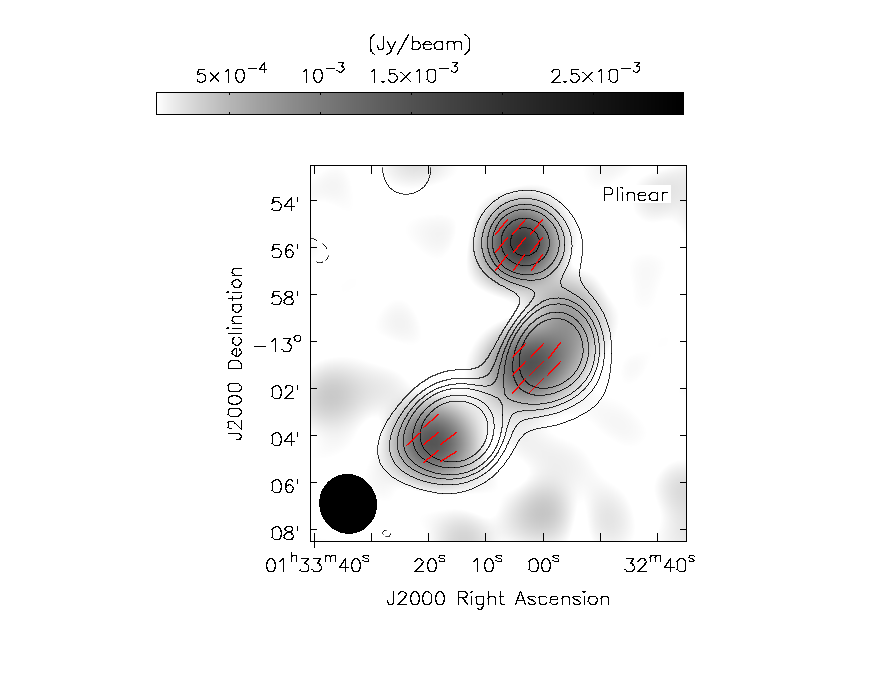}
\end{center}
\caption{The figure shows the linear polarization image in grey scale overlay with the total intensity I contours. The contours are at $[-1, 1, 2, 3, 4, 6, 8] \times 3\sigma$ mJy/beam, where $\sigma = 2$ in units of mJy/beam.
The black circle in the right corner shows the synthesized beam of the KAT-7. The red dashes show the polarization electric vectors rotated by 90 deg to show the magnetic field direction.  The polarization angle was derived where the  polarization intensity exceeds $10\sigma_{rms}$ where $\sigma_{rms}=0.1$ mJy/beam.}
\label{fig:GRG_plin}
\end{figure}

Figure \ref{fig:GRG_plin} shows the  linear polarization image in grey scale overlay with the total intensity I contours. The contours are at $[-1, 1, 2, 3, 4, 6, 8] \times 3\sigma$ mJy/beam, where $\sigma = 2$ in units of mJy/beam. The red dashes show the polarization electric vectors rotated by 90 deg to show the magnetic field direction. The polarization angle was derived where the  polarization intensity exceeds $10 \times \sigma_{rms}$ where $\sigma_{rms}=0.1$ mJy/beam.
We notice the coherent linear polarization orientation in the NW lobe and in the SE lobe 
(actually a combination of the SE lobe and the GRG core polarization).

The polarization calibration was carried out using 3C138.  We used the procedures as described by Riseley et al. (2015).  Cross hand delays and phases were determined assuming a non zero polarization model. Stokes parameters were derived and compared to Perley \& Butler (2013). From these data we recovered a fractional  polarization of $8.52 \pm 0.55 \%$ and a mean electric vector position angle of $-9.87 \pm 0.68$ deg.
These values are consistent with those of other giant radio galaxies  as derived by Machalski et al. 
(2006) and Mack et al. (1997).

We also find evidence in this image of a separate polarized source about 7$^{\prime}$ North of the GRG
(see Fig. \ref{fig:GRG_plin}). 
This compact radio source is likely associated with a WISE source at coordinates RA=$01^h 33^m 03^s$, DEC=$-12^\circ  55^{\prime} 53^{\prime \prime}$ (J2000.0), that is however confused.  
The source J013303.19-125553.5 (magnitude W1$=15.94 \pm 0.05$) may also be extragalactic based on its colors, but it is highly confused by three foreground stars, and hence it is not possible to estimate a reliable photometric redshift for this source.
However, based on its colors it could be another member of a large-scale structure around the GRG: in fact, in the vicinity (at a distance of order of 1-2$^{\prime}$) other similar WISE color sources (hence likely extragalactic sources) are present, suggesting the possible association of these radio sources with a galaxy cluster over-density. 

The polarized source J013303.19-125553.5 looks like a high-polarization compact source suggesting also the possibility that it is a blazar-like source.
More follow-up work is, however, needed to confirm the nature of this source.

\begin{figure}
\centering
\includegraphics[scale=0.34]{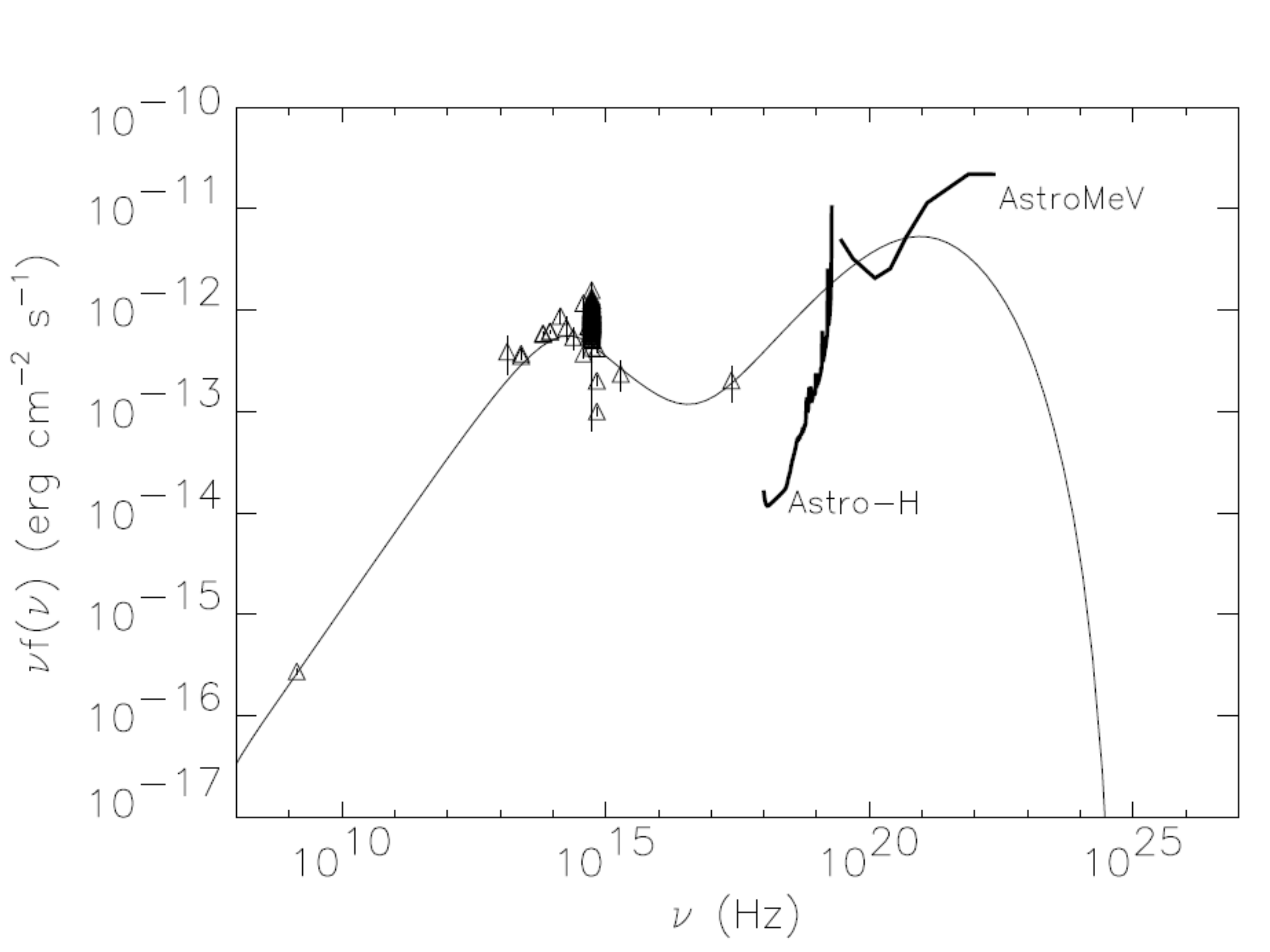}
\caption{The GRG core SED fitted with an one-zone SSC model with the parameters in Table \ref{tab.param.fit}.
We show also the sensitivity curves for ASTRO-H with 100 ks of time integration (from http://astro-h.isas.jaxa.jp/researchers/sim/sensitivity.html) and for AstroMeV for 5 yrs of survey (from http://astromev.in2p3.fr/sites/default/files/Takahashi-COCOTE\_sens.jpg).
SED data are from NVSS, WISE W1-W4, 2MASS, CATALINA Real-time Transient Survey (RTS), United States Naval Observatory B1.0 Catalog (USNO B1),  GALEX all-sky survey (AIS) far ultra-violet (FUV) band, and RASS.
}
\label{fig:fit8}
\end{figure}

In order to validate the physical nature of the central radio galaxy core J013313.50-130330.5, we show in Fig.\ref{fig:fit8} the multi-frequency SED of the object associated to the radio/mid-IR/X-ray source S2. The SED in the $log(\nu) - log(\nu f(\nu))$ plane shows a synchrotron peak at $\nu \approx 10^{14}$ Hz consistent with the SED of a radio galaxy, blazar-like core. 
A simple 1-blob synchrotron self-Compton (SSC) model (see, e.g., Jones, O'dell \& Stein 1974, see also Colafrancesco, Marchegiani \& Giommi 2010 for a recent application of our specific SSC model) with a double power law electron spectrum is able to reproduce the multifrequency SED with the parameters given in Table \ref{tab.param.fit}. In the fitting procedure for the SED, we have fixed the radio galaxy redshift to the photometric one $z = 0.3$ estimated with infrared data, and we chose a small value of the boosting parameter $\delta=2$  where
\begin{equation}
\delta=\frac{1}{\Gamma (1-\beta \cos \theta)} \; ,
\label{param.boosting}
\end{equation}
with $\beta=v/c$ and $\Gamma=(1-\beta^2)^{-1/2}$, in terms of the blob velocity $v$ and of the viewing angle $\theta$, so that this parameter is compatible with the typical one of a misaligned AGN. 
We chose also a reference value for the jet magnetic field of $10$ G as it is found in the inner cores of other radiogalaxies  like, e.g., M87, with a magnetic field in the range $\sim 1-15$ G (see Kino et al. 2014), or PKS1830-211, with a magnetic field of tens G or even higher (see Mart\'i-Vidal et al. 2015), and leave the other parameters free to vary in order to find the best possible combination of them that fit the SED. The fit shown in Fig.\ref{fig:fit8} is good at reproducing the observed data from the radio, to the IR/optical/UV and X-rays bands. 
Based on the best-fit parameters of the SED of the radio galaxy core with the value $\delta=2$, one can show that the radio galaxy jet has a maximum inclination of $\theta \approx 30$ deg with respect to the line of sight, a value obtained for $\delta = \Gamma$ in eq.(\ref{param.boosting}), while the minimum value of $\theta$ is not constrained due to the degeneracy existing with the value of $\Gamma$.
This specific model does not predict a $\gamma$-ray emission detectable with the Fermi-LAT or Cherenkov telescopes like HESS because the maximum of the IC branch falls in the spectral region around 1--10 MeV. Observations in this energy band would be hence desirable to test the predictions of our model or to further assess the visibility of this object in the GeV/TeV energy range.  The proposed AstroMeV space experiment (see http://astromev.in2p3.fr/) operating in the 0.1--100 MeV range has the sensitivity to detect such an object and provide a spectrum in the MeV region (see Fig.\ref{fig:fit8}).
We also notice that the future ASTRO-H space experiment (see http://astro-h.isas.jaxa.jp/en/) has the sensitivity to perform a spectral analysis of the IC branch of the SED of the core of the GRG as well as of the IC X-ray emission expected from the GRG lobes that will clarify their structure and origin. 
Further observations of the optical spectrum will be important to assess the nature of this object and of the host galaxy and to obtain an accurate spectroscopic redshift.

\begin{table*}
\centering
 \caption{Parameters of the SSC model in Fig.\ref{fig:fit8}. Col.1: central electron density; Col.2 and 3: spectral indices of the double power law electron spectrum at low and high energies; Col.4: electron Lorentz factor where the spectrum changes the slope; Col.5: central magnetic field; Col.6: boosting factor (see eq.\ref{param.boosting}); Col.7: radius of the emitting region; Col.8: redshift.}
 \label{tab.param.fit}
\begin{tabular}{*{8}{c}}
\hline
\hline
 $\log N_0(\mbox{cm}^{-3})$ & $p_1$ & $p_2$ & $\log \gamma_b$ & $B_0$ (G) & $\delta$ & $r$ (pc) & $z$\\
\hline
5.3 & 1.5 & 3.9 & 3.2 & 10 & 2 & 0.0005 & 0.3\\
\hline
 \end{tabular}
 \end{table*}

\section{Discussion and conclusions}
\label{sec:disconc}

The discovery of an extended and highly aligned radio emission in the source centered at RA$=01^h 33^m 13.5^s$ and DEC$=-13^\circ  03^{\prime} 31^{\prime \prime}$ (J2000) has been obtained with KAT-7 and suggests the existence of a GRG whose larger radio lobe extends on angular scales of $\ga 4^{\prime}$, corresponding to a linear distance of more than $\ga 1$ Mpc at the photometric redshift $z \approx 0.3$ of the source.
This GRG is not in the list of the newly detected giant radio galaxies from NVSS radio survey at 1.4 GHz conducted by Solovyov \& Verkhodanov (2014). Their automatic search procedure selected, in fact, only sources larger than 4$^{\prime}$. Therefore our newly detected GRG  was missed since the NVSS contours extension for each one of the lobes for this source are smaller ($\sim 3^{\prime}$). This emphasizes the relevance of KAT-7 in the study of these sources because the small baseline configuration of the KAT-7 is able to detect more extended radio emission than the NVSS and yet being consistent and encompassing the spatial distribution of the NVSS radio sources (see Fig. \ref{fig:GRG_nvss}).\\
The multifrequency study of this extended radio source supports the discovery of a suspected GRG whose core is likely hosted by the WISE object J013313.50-130330.5 which is associated with the NVSS source S2 and the ROSAT X-ray source 1RXSJ013313.8-13031.
The analysis of the multi-frequency SED of the source J013313.50-130330.5 indicates that  the emitting region appears to be very small, suggesting that the source of the emission is very compact and compatible with a central AGN, from where  giant radio lobes can originate, as in many other cases of misaligned AGNs like FR-II objects.
The geometry of this source is consistent with the extended jets of a radiogalaxy inclined by at most 30 deg with respect to the line of sight and aligned along the SE to NW direction on the sky.
The extension and the structure of this source suggests the possible discovery of a new GRG emitting two symmetric powerful jets/lobes  powered by the strong nuclear activity of the source J013313.50-130330.5.
At the photometric redshift of $\approx 0.3$, this radio source would have a core luminosity of $P_{1.4 GHz} \approx 5.52 \times 10^{24}$ W Hz$^{-1}$, and luminosities $P_{1.4 GHz} \approx 1.29 \times 10^{25}$ W Hz$^{-1}$ for the brightest NW lobe and $P_{1.4 GHz} \approx 0.46 \times 10^{25}$ W Hz$^{-1}$ for the SE lobe (these luminosities have been estimated from the NVSS fluxes at 1.4 GHz), consistent with the typical luminosities of GRG (see, e.g., Malarecki et al. 2013).\\
High sensitivity radio observations of this GRG are needed to reveal the existence of the collimated radio jet linking the central core to the extended lobes. Further high-resolution observations with the  Karl G. Jansky Very Large Array (VLA), and possibly also with the Very Long Baseline Array (VLBA),  would be able to properly analyze the spatial structure of this source and to determine the core to total radio power ratio, its brightness temperature, the jet velocity and its orientation.

The suspected GRG discovery with KAT-7 was possible due to the large field of view of KAT-7 and its sensitivity to extended and low-surface brightness radio emission, indicating that these GRGs (FR-II systems) can be more common that what is believed based on NVSS and/or FIRST surveys. 
The discovery presented herein is, therefore, suggesting a potential science case for MeerKAT and ASKAP-EMU, and ultimately for the SKA, in that it has the capability to discriminate between models for the cosmological evolution of misaligned AGNs and the evolutionary dichotomy of FR-II vs. FR-I radiogalaxies.

\section*{Acknowledgments}
SC and THJ acknowledge support by the South African Research Chair Initiative of the Department of Science and Technology and National Research Foundation and by the Square Kilometre Array. NM and PM acknowledge support by the DST/NRF SKA post-graduate bursary initiative.
We thank the Referee for his/her several constructive comments on our paper.

\label{lastpage}


\begin{thebibliography}{}
\bibitem[\protect\citeauthoryear{Armstrong et al.}{2013}]{Armstrong2013}
Armstrong R.P. et al., 2013, MNRAS, 433, 1951
\bibitem[\protect\citeauthoryear{Brown et al.}{2014}]{Brown2014}
Brown M.J.I. et al., 2014, ApJS, 212, 18
\bibitem[\protect\citeauthoryear{Butenko et al.}{2014}]{Butenko2014}
Butenko A.V., Dagkesamanskii R.D., Samodurov V.A., Tyulbashev S.A., 2014, Astron. Rep., 58, 363
\bibitem[\protect\citeauthoryear{Carignan et al.}{2013}]{Carignan2013}
Carignan C., Frank B.S., Hess K.M., Lucero D.M., Randriamampandry T.H., Goedhart S., Passmoor S.S., 2013, AJ, 146, 48
\bibitem[\protect\citeauthoryear{Colafrancesco 2008}{2010}]{Colafrancesco2008}
Colafrancesco S., 2008, MNRAS, 385, 2041
\bibitem[\protect\citeauthoryear{Colafrancesco \& Marchegiani}{2011}]{CM2011}
Colafrancesco S., Marchegiani P., 2011, A\&A, 535, 108
\bibitem[\protect\citeauthoryear{Colafrancesco, Marchegiani \& Giommi}{2010}]{CMG2010}
Colafrancesco S., Marchegiani P., Giommi P., 2010, A\&A, 519, 82
\bibitem[\protect\citeauthoryear{Colafrancesco et al.}{2014a}]{Col14a}
Colafrancesco S., Mhlahlo N., Oozeer N., 2015a, A\&A, submitted   
\bibitem[\protect\citeauthoryear{Colafrancesco et al.}{2014b}]{Col14b}
Colafrancesco S., Mhlahlo N., Oozeer N., 2015b, A\&A, submitted  
\bibitem[\protect\citeauthoryear{Ishwara-Chandra \& Saikia}{1999}]{Ishwara1999}
Ishwara-Chandra C.H., Saikia, D.J., 1999, MNRAS, 309, 100
\bibitem[\protect\citeauthoryear{Ishwara-Chandra \& Saikia}{2002}]{Ishwara2002}
Ishwara-Chandra C.H., Saikia D.J., 2002, New Astron. Rev., 46, 71
\bibitem[\protect\citeauthoryear{Jarrett et al.}{2011}]{Jarrett2011}
Jarrett T.H. et al., 2011, ApJ, 735, 112
\bibitem[\protect\citeauthoryear{Jones et al.}{1974}]{Jonesetal1974}
Jones T.W., O'dell S.L., Stein W.A., 1974, ApJ, 188, 353
\bibitem[\protect\citeauthoryear{Kino et al.}{2014}]{Kino2014}
Kino M., Takahara F., Hada K., Doi A., 2014, ApJ, 786, 5
\bibitem[\protect\citeauthoryear{Kronberg et al.}{2004}]{Kronberg}
Kronberg P.P., Colgate S.A., Li H., Dufton Q.W., 2004, ApJ, 604, L77
\bibitem[\protect\citeauthoryear{Lara et al.}{2001}]{Lara2001}
Lara L., Marquez I., Cotton W.D., Feretti L., Giovannini G., Marcaide J.M., Venturi T., 2001, A\&A, 378, 826
\bibitem[\protect\citeauthoryear{Machalski et al.}{2001}]{Machalski2001}
Machalski J., Jamrozy M., Zola S., 2001, A\&A, 371, 445
\bibitem[\protect\citeauthoryear{Machalski et al.}{2006}]{Machalski2006}
Machalski J., Jamrozy M., Zola S., Koziel D., 2006, A\&A, 454, 85
\bibitem[\protect\citeauthoryear{Mack et al.}{1997}]{Mack1997}
Mack K.-H., Klein U., O'Dea C.P., Willis A.G., 1997, A\&AS, 123, 423 
\bibitem[\protect\citeauthoryear{Magliocchettiet al.}{1998}]{Magliocchetti1998}
Magliocchetti M., Maddox S.J., Lahav O., Wall J.V., 1998, MNRAS, 300, 257
\bibitem[\protect\citeauthoryear{Malarecki et al.}{2013}]{Malarecki2013}
Malarecki J.M., Staveley-Smith L., Saripalli L., Subrahmanyan R., Jones D.H., Duffy A.R., Rioja M., 2013, MNRAS, 432, 200
\bibitem[\protect\citeauthoryear{Mart-Vidal et al.}{2015}]{Marti2015}
Mart\'i-Vidal I., Muller S., Vlemmings W., Horellou C., Aalto S., 2015, Science, 348, 311
\bibitem[\protect\citeauthoryear{Norris et al.}{2009}]{Norrisetal2007}
Norris R.P., the EMU team, 2009, in Heald G., Serra P., eds, Proc. of Sci., PRA2009, Panoramic Radio Astronomy: Wide-field 1-2 GHz research on galaxy evolution, id.33
\bibitem[\protect\citeauthoryear{Perley}{2013}]{Perley2013}
Perley R.A., Butler B.J., 2013, ApJS, 206, 16
\bibitem[\protect\citeauthoryear{Riseley et al.}{2015}]{Riseleyetal2015}
Riseley C.J., Scaife A.M.M., Oozeer N., Magnus L., Wise M.W., 2015, MNRAS, 447, 1895
\bibitem[\protect\citeauthoryear{Saripalli et al.}{2005}]{Saripallietal2007}
Saripalli L., Hunstead R.W., Subrahmanyan R., Boyce E., 2005, AJ 130, 896
\bibitem[\protect\citeauthoryear{Scaife et al.}{2015}]{Scafeetal2015}
Scaife A.M.M., Oozeer N., de Gasperin F., Brüggen M., Tasse C., Magnus L., 2015, MNRAS, 451, 4021
\bibitem[\protect\citeauthoryear{Schoenmakers et al.}{2001}]{Schoenmakersetal2001}
Schoenmakers A.P., de Bruyn A.G., R\"ottgering H.J.A., van der Laan H., 2001, A\&A, 374, 861
\bibitem[\protect\citeauthoryear{Silva et al.}{1998}]{Silvaetal1998}
Silva L., Granato G.L., Bressan A., Danese L., 1998, ApJ, 509, 103
\bibitem[\protect\citeauthoryear{Solovyov \& Verkhodanov}{2014}]{Solovyov2014}	
Solovyov D.I., Verkhodanov O.V., 2014, Astrophys. Bull., 69, 141
\bibitem[\protect\citeauthoryear{Subrahmanyan et al.}{2008}]{Subrahmanyan2008}	
Subrahmanyan R., Saripalli L., Safouris V., Hunstead R.W., 2008, ApJ, 677, 63
\bibitem[\protect\citeauthoryear{Voges et al.}{1999}]{RASS}	
Voges W. et al., 1999, A\&A, 349, 389
\end{thebibliography}
\end{document}